\pdfoutput=1
\documentclass[10pt, journal, compsoc, twoside, hidelinks]{IEEEtran}

\usepackage[utf8]{inputenc}
\usepackage[T1]{fontenc}
\usepackage[english]{babel}

\usepackage[nocompress]{cite}
\newcommand{\ignore}[1]{}
\usepackage{fancyhdr}
\usepackage{booktabs}
\usepackage{xfrac}
\usepackage{algorithm}
\usepackage{algorithmic}
\usepackage{xspace}
\usepackage{tabularx}
\usepackage{placeins}
\usepackage[para]{threeparttable}
\usepackage{multirow}
\usepackage{multicol}
\usepackage{color}
\usepackage{tikz}
\usepackage{amssymb}
\usepackage{pifont}
\usepackage{colortbl}
\usepackage{amsmath}
\usepackage{listings}
\usepackage{etoolbox}
\usepackage[printwatermark]{xwatermark}
\usepackage{mwe}
\usepackage[shortcuts]{extdash}
\usepackage{xspace}
\usepackage{tabularx}
\usepackage{booktabs}
\usepackage{multirow}
\usepackage{tikz}
\usepackage{enumitem}
\usepackage{graphicx}
\usepackage{verbatim}
\usepackage{latexsym}
\usepackage{setspace}
\usepackage{footnote}
\usepackage{textgreek}
\usepackage{acronym}
\usepackage{ulem}

\newcommand{\tabref}[1]{Table~\ref{#1}}
\newcommand{\figref}[1]{Fig.~\ref{#1}}

\newcommand{\io}{I/O}

\newcommand{\effectiveArea}{5.11mm\textsuperscript{2}}

\newcommand{\socareanoefpga}{1.11mm\textsuperscript{2}}

\newcommand{\risckyPMPovh}{14\%}

\newcommand{\FPGAAREAspipulpino}{20\%}
\newcommand{\FPGAAREAmuntaniala}{10\%}
\newcommand{\FPGAAREAbnnacc}{42\%}
\newcommand{\maxFLL}{2.1\,GHz}

\newcommand{\FPGAAREAcrcacc}{2\%}

\acrodef{ASIC}{Application specific integrated circuit}
\acrodef{FPGA}{Field Programmable Gate Array}
\acrodef{MCU}{Microcontroller unit}
\acrodef{SoC}{System-On-Chip}
\acrodef{CPU}{Central Processing Unit}
\acrodef{API}{Application Programming Interface}
\acrodef{FLL}{Frequency-locked loop}

\acrodefplural{ASIC}[ASICs]{Application specific integrated circuits}
\acrodefplural{FPGA}[FPGAs]{Field Programmable Gate Arrays}
\acrodefplural{MCU}[MCUs]{Microcontroller units}
\acrodefplural{SOC}[SoCs]{System-On-Chips}
\acrodefplural{API}[APIs]{Application Programming Interfaces}
\acrodefplural{FLL}[FLLs]{Frequency-locked loops}

\acrodef{NTC}{Near-Threshold Computing}
\acrodef{FDSOI}{Fully Depleted   Silicon-On-Insulator}
\acrodef{ADC}{Analog-to-Digital converter}
\acrodef{AFE}{analog front-end}
\acrodef{DSP}{digital signal processor}
\acrodef{AP}{Action Potential}
\acrodef{EEG}{Electroencephalography}
\acrodef{LFP}{Local Field Potential}
\acrodef{SoA}{\textit{state-of-the-art}}
\acrodef{PPA}{Power-Performance-Area}
\acrodef{DWT}{\textit{Discrete Wavelet Transform}}
\acrodef{ISA}{Instruction Set Architecture}
\acrodef{PMP}{Physical Memory Protection}
\acrodef{ROM}{read-only-memory}
\acrodef{SRAM}{static random-access-memory}
\acrodef{APB}{AMBA Advanced Peripheral Bus}
\acrodef{GF22FDX}{Globalfoundries GF22FDX}
\acrodef{MAC}{Multiply-and-Accumulate}
\acrodef{FCB}{Fabric Configuration Block}
\acrodef{OTS}{Off-the-Shelf}
\begin{document}

\title{Arnold: an eFPGA-Augmented RISC-V SoC for Flexible and Low-Power IoT End-Nodes}

\makeatletter
\patchcmd{\@maketitle}
  {\addvspace{0.5\baselineskip}\egroup}
  {\addvspace{-4\baselineskip}\egroup}
  {}
  {}
\makeatother

\author{Pasquale Davide Schiavone, Davide Rossi \IEEEmembership{Member, IEEE}, Alfio Di Mauro, Frank G\"urkaynak, Timothy Saxe, Mao Wang, Ket Chong Yap, Luca Benini \IEEEmembership{Fellow, IEEE}
\thanks{P. D. Schiavone, A. Di Mauro, F. G\"urkaynak, and L. Benini are with the Integrated Systems Laboratory, D-ITET, ETH Z\"urich, 8092 Z\"urich, Switzerland.
D. Rossi is with the Energy-Efficient Embedded Systems Laboratory, DEI, University of Bologna, 40126 Bologna, Italy.
T. Saxe, M. Wang, and K.C. Yap are with the QuickLogic Corporation, 2220 Lundy Ave, San Jose, CA 95131, United States of America.
 }}

\maketitle

\begin{abstract}
A wide range of Internet of Things (IoT) applications require powerful, energy-efficient and flexible end-nodes to acquire data from multiple sources, process and distill the sensed data through near-sensor data analytics algorithms, and transmit it wirelessly. This work presents \textit{Arnold}: a 0.5\,V to 0.8\,V, 46.83\,\textmu W/MHz, 600\,MOPS fully programmable RISC-V \ac{MCU} fabricated in 22\,nm \ac{GF22FDX} technology, coupled with a \ac{SoA} microcontroller to an embedded \ac{FPGA}. We demonstrate the flexibility of the \ac{SoC} to tackle the challenges of many emerging IoT applications, such as \textit{(i)} interfacing sensors and accelerators with non-standard interfaces, \textit{(ii)} performing on-the-fly pre-processing tasks on data streamed from peripherals, and \textit{(iii)} accelerating near-sensor analytics, encryption, and machine learning tasks. A unique feature of the proposed \ac{SoC} is the exploitation of body-biasing to reduce leakage power of the embedded \ac{FPGA} (eFPGA) fabric by up to 18$\times$ at 0.5\,V, achieving \ac{SoA} state bitstream-retentive sleep power for the eFPGA fabric, as low as 20.5\,\textmu W. The proposed \ac{SoC} provides 3.4$\times$ better performance and 2.9$\times$ better energy efficiency than other fabricated heterogeneous re-configurable \acp{SoC} of the same class.

\end{abstract}


\section{Introduction}
\label{sec:introduction}

The end-nodes of the IoT require energy-efficient, powerful, and flexible ultra-low-power computing platforms to deal with a wide range of near-sensor applications \cite{bonomi2012fog}. These \acp{SoC} must be able to connect to low-power sensors such as arrays of microphones \cite{8648460}, cameras \cite{8715489}, electrodes to monitor physiological activities \cite{8662605}, to analyze and compress data using advanced algorithms, and transmit them wirelessly over the network. Signal processing algorithms are executed in such devices to reduce complex raw data to simple classifications tags that classify data, to extract only relevant information (e.g., \cite{Villegas}), or to filter, encrypt, anonymize data. Compressing and distilling information that travels from IoT devices to the cloud, brings multiple benefits in power, performance, and bandwidth across the whole IoT infrastructure.

Depending on the constraints of the application such as flexibility, performance, power, and cost, IoT computing platforms can be implemented as hardwired \acp{ASIC}, programmable hardware (or soft-hardware) on \acp{FPGA}, or as software programmable on \acp{MCU}. Hardwired, fixed-function \acp{ASIC} offer the best energy and energy efficiency, but they lack versatility and require long time-to-market \cite{rossi2013multicore}. Hence, their usage is preferred in highly standardized applications or specialized single-function products.

On the other side of the spectrum, \acp{MCU} are the de-facto standard platforms for IoT applications thanks to their high versatility, low-power, and low-cost. \ac{SoA} \acp{MCU} can offer competitive \ac{PPA} figures by leveraging parallel \ac{NTC} \cite{dreslinski2010near}, and advanced low-power technologies such as \ac{FDSOI} coupled with performance-power management techniques such as body-bias \cite{8065010} and power-saving states \cite{8715500}. As it has been shown in \cite{SleepWalker, rossi2017energy, Bol19, 8715500}, these techniques make possible the use of \acp{MCU} on edge computing devices, meeting \ac{PPA} constraints for a wide range of applications in the IoT domain, yet providing high versatility. To increase performance, \acp{MCU} are often customized with on-chip full-custom accelerators that speed up the execution of part of the applications as for example neural-networks \cite{ContiBNN}, frequency-domain-transforms \cite{bansal2019closely}, linear algebra \cite{Cavalcante}, security engines \cite{fritzmann2019towards}. The resulting heterogeneous system has thus both the flexibility of \acp{MCU}, and competitive performance and efficiency of hardwired \acp{ASIC} on specific domains.

\acp{FPGA} fill the gap between \acp{ASIC} and \acp{MCU} as they offer versatility via hardware programmability (which usually needs longer design and verification time than software), and they allow exploiting spatial computations typical of \acp{ASIC} designs, as opposed to sequential execution. For these reasons, \acp{FPGA} are used in a wide range of applications, from machine learning \cite{8419763, 8892142, 8439040}, sorting \cite{8735541}, and cryptography accelerators for data centers \cite{sawant2020spartan}, to smart instruments \cite{8542775}, \textcolor{black}{analog-to-digital converters} \cite{7593301}, to low-power systems for wearable applications \cite{8119727}, control-logic systems \cite{8464146}, and for implementing smart-peripherals connected to \acp{SoC} \cite{6119933, Williams}.

Increased integration density of modern \acp{SoC} allowed a reasonably sized FPGA array to be integrated as part of an on-chip system. Such embedded FPGAs (eFPGAs) are used to enable post-silicon soft-hardware programmable functions in \acp{SoC} or \acp{MCU} to make updates on accelerators or custom peripherals.
As for the FPGA case, hardwired accelerators or peripherals outperform their eFPGA-based implementations, but lack flexibility and post-fabrication reconfigurability.
The benefit of integrating eFPGAs into SoCs is the possibility to increase performance by specializing the \acp{SoC} for one particular domain that can change over time, increasing the product life-time and application span.

In this paper, we present \textit{Arnold}: a RISC-V based MCU extended with an eFPGA, implemented in \ac{GF22FDX} technology.
The contribution of the presented heterogeneous SoC design and silicon demonstrator are summarized as follows.

\begin{enumerate}
   \item \emph{Architectural Flexibility}: to enable architectural flexibility that fully exploits the configurable logic. The eFPGA is connected with the rest of the system with different interface options on the data-plane: \textit{i}) a direct connection to the \io{} DMA engine on the SoC - to process and filter data streams on their way from/to on-chip shared memory buffers in memory; \textit{ii}) a high-bandwidth, low-latency interface to the memory of the RISC-V core - to interleave with zero-copy FPGA-accelerated parallel processing and sequential processing by the core; \textit{iii}) a direct GPIO interface to implement master or slave peripheral ports for non-standard off-chip digital sensors or actuators. On the control plane we provide: \textit{i}) an \ac{APB} interface to allow the user to configure the mapped soft-hardware; \textit{ii}) sixteen interrupts to notify the CPU.

    \item \emph{Power Management}: thanks to reverse body-bias (RBB) enabled by conventional-well \ac{FDSOI} technology used for the physical implementation of the eFPGA fabric, leakage power can be reduced by 18x to 20.5\,\textmu W (featuring a fully state retentive bitstream) when eFPGA functionality is not required.

    \item \emph{Leading Edge Performance and Energy Efficiency}: the \ac{SoC} achieves SoA performance and efficiency, leveraging a voltage and frequency scalable architecture from 0.5\,V to 0.8\,V, with a peak energy efficiency of 46.83\,\textmu W/MHz at 0.52\,V and a maximum frequency of 600\,MHz at 0.8\,V. The proposed SoC achieves 3.4$\times$ better performance and 2.9$\times$ better energy efficiency than \ac{SoA} \acp{MCU} augmented with eFPGA built for the same power target applications \cite{Borgatti, Lodi, Renzini}.
\end{enumerate}

The remainder of the paper is organized as follows: Section II provides a review of related works. In Section III, the architecture of the proposed SoC is described, including all its components. In Section IV and V, the software and tools for the proposed SoC, its physical design, and silicon measurements are described respectively, whereas, in Section VI, use cases for the proposed work are reported as application examples. The paper concludes in Section VII.

\section{Related Work}
\label{sec:related}

In this section, we review devices that define the boundaries of its design space: \acp{MCU}, \acp{FPGA}, eFPGAs, and heterogeneous reconfigurable SoCs.

\begin{table*}
    \centering
    \begin{threeparttable}
    \label{tab:relatedwork}
    \caption{Summary of related work: (left) MCUs programmable via Software and their accelerators; (center) FPGAs programmable via Soft-Hardware design; (right) eFPGAs programmable via Soft-Hardware design. }

    \begin{tabular}{@{}rlcrlcrl@{}}
        \toprule
        \multicolumn{2}{c}{\textbf{MCU}} &
         &
        \multicolumn{2}{c}{\textbf{FPGA}} &
         &
        \multicolumn{2}{c}{\textbf{eFPGA}}
        \\
        \midrule
        \textbf{Single Core}&
        \cite{Bol19, NXPiMX, STmicro, schiavone2018quentin}& &
        \textbf{Low Power}&
        \cite{igloonano, latticeice}& &
        \textbf{StandAlone}&
        \cite{menta, achronix, flexlogix, quicklogic}
        \\
        \textbf{SW Accelerator}&
        \cite{NXPiMX7ULP, 8715500}& &
        \textbf{Low Power SoC}&
        \cite{SmartFusion}& &
        \textbf{MCU SoC}&
        \cite{Borgatti, Lodi, Renzini},
         \\
        \textbf{HW Accelerator}&
        \cite{ContiBNN, SiLabs}& &
        \textbf{HP}&
        \cite{xilinxVirtex, intelCyclone}& &
        &
        \textbf{This Work}
         \\
        \textbf{HW/SW Accelerator}&
        \cite{GAP8, 7927716}& &
        \textbf{HP SoC}&
        \cite{Xilinx, ArriaV}& &
        \textbf{HP SoC}&
        \cite{Whatmough}
        \\
    \bottomrule
    \end{tabular}

    \end{threeparttable}
    \vspace{-4mm}
\end{table*}

\subsection{\acp{MCU}}
\label{subsec:mcu}

In the context of edge-computing systems, \acp{MCU} need to provide significant performance within a limited power budget, and the flexibility needed to cope with a wide variety of sensors and algorithms. Most \ac{OTS} \acp{MCU} use energy-efficient \acp{CPU} based on ARM Cortex-M family of cores, such as the NXP i.MXRT1050 \cite{NXPiMX}, the STMicroelectronics STM32L476xx family \cite{STmicro}, or the Silicon Labs EFM32 Giant Gecko 11 \cite{SiLabs}, all featuring a power budget within a few tens of mW. To interface with a large variety of external devices, these systems offer a wide set of peripherals such as I2C, UART, SPI, and GPIOs. \acp{SoA} energy-efficient \acp{MCU} optimized for ultra-low-power (3\textmu W/MHz) \cite{Bol19} and performance (938\,MHz) \cite{schiavone2018quentin} have been implemented in \ac{FDSOI} technology leveraging body-biasing to compensate process-voltage-temperature (PVT) variations, and to control performance and power to achieve higher energy efficiency.

Although software provides high versatility, some applications still need performance that a single CPU cannot deliver. For this reason, several \acp{MCU} are extended with custom accelerators, for example, the binary neural-network accelerator presented in \cite{ContiBNN} or the cryptography engine integrated into \cite{SiLabs}. To improve flexibility with respect to dedicated accelerators, there are \acp{MCU}  that combine multiple heterogeneous CPUs managing different tasks, for example, the NXP i.MX 7ULP Applications Processor \cite{NXPiMX7ULP}, which combines an application ARM processor (ARM Cortex-A7) with a real-time CPU (ARM Cortex-M4) for performance and power trades off. Other approaches leverage parallel clusters of processors to improve the energy efficiency of near sensor analytics workloads, such as Mr.Wolf \cite{8715500}, featuring an 8-core cluster based on DSP-enhanced RISC-V cores controlled by a smaller core managing the \io{}s, the runtime, and SoC control functions. These systems can choose to divide the workload as a subset of processors to meet the performance target at the lowest energy budget \cite{SchiavonePatmos}. Finally, heterogeneous systems like GAP-8 from GreenWaves Technologies \cite{GAP8} and Fulmine \cite{7927716}, combine both custom and parallel software programmable accelerators providing a step forward for performance and flexibility of embedded platforms for signal processing. \textcolor{black}{Although these platforms are compelling and flexible to run signal processing tasks for typical end-nodes, they are less efficient than reconfigurable devices such as \acp{FPGA} when dealing with non-standard sensors.}

\subsection{\acp{FPGA}}

\acp{FPGA} are reconfigurable devices that on one hand can exploit spatial computations typical of ASIC designs but still retain the capability of being reconfigured after fabrication. They range from high-end FPGAs used for acceleration of high-performance workloads to ultra-low-power, small and low-cost technology implementations, as discussed further in this section.

High-end FPGAs, such as the Xilinx Virtex Ultrascale devices \cite{xilinxVirtex} and the Intel Cyclone 10 GX device \cite{intelCyclone}, have millions of LUTs, flip-flops, DSP-blocks, and SRAM macros containing Mbytes of memory.
To extended their capabilities in the embedded application domain running software, such FPGAs are often programmed with soft-CPUs \cite{ng2016soft}.
The users can implement a deeply pipelined core with multiple issues to achieve high performance, or a tiny soft-core with a small area footprint for control applications \cite{heinz2019catalog, holler2019open}, and offload part of the control functionalities executed in SW to the soft-CPU.
For example, Choi et al.  \cite{5418887} presented a FPGA-based 20\,k-Word speech recognizer using a Xilinx Virtex-4 FPGA where the computationally less demanding tasks are executed in SW, whereas the rest of the algorithms is accelerated in HW.

As soft-cores are limited in performance \cite{7776886} and occupy resources, FPGAs are often extended with hard-CPUs as application processors (usually ARM-based embedded processors such as the Xilinx Zynq-7000 SoC \cite{Xilinx} and Intel Arria V SoC \cite{ArriaV}, in the case of Microsemi PolarFire \cite{polarfire} RISC-V processors).
As a result, high-end FPGAs have typical power consumption in the order of tens of Watts \cite{pandey2017performance}, and they are usually used as high-performance accelerators on servers connected via Ethernet or PCI interfaces \cite{8735540}.

In the low-power domain, FPGAs are typically realized with a less aggressive process than high-end FPGAs. They are usually smaller, cheaper, and as a result, have lower performance than the others. Examples are the Microsemi IGLOO nano \cite{igloonano}, which has up to 3\,k logic elements\footnote{One logic element is composed of one 4-input LUT and one flip-flop}, or the Lattice Semiconductor iCE40 UltraLite \cite{latticeice}, which has more than 1K of LUTs+flip-flops. Both consume from a few \textmu W to hundreds of mW. These FPGAs are used to extend the \io{} subsystem of embedded controllers \cite{8060447}, even with simple data pre-processing engines to lower the bandwidth coming from sensors \cite{7348330, 8119727}.
In the low-end space, FPGAs can also be extended with CPUs to leverage HW/SW co-designed IoT nodes.
An industrial RISC-V based soft-core is provided by the Microsemi Mi-V RV32, ready to be integrated into the SmartFusion2 SoC \cite{SmartFusion} or in the IGLOO FPGA \cite{miv} in an area footprint of 10\,k-26\,k LEs. Other RISC-V based solutions have emerged during the \textit{RISC-V SoftCPU Contest} in December 2018, with the VexRiscv soft-core as the winner.
Hard-CPUs are also used as in the Microsemi SmartFusion2 SoC in 65nm \cite{SmartFusion}, which proposes an MCU-class (ARM Cortex-M) core running at 166\,MHz and an FPGA with DSP blocks and up to 150\,k logic elements, 656\,kB\footnote{512\,Bytes of Non-Volatile Memory} of memory, and power consumption in the order of hundreds of mWatts. Examples that use the Microsemi SmartFusion2 SoC can be found in Gomes at al. \cite{Gomes}, which proposes a system where most of the tasks are executed by the ARM core, whereas the FPGA is used for accelerating critical network kernels. In Fournaris et al. \cite{Fournaris}, the operating system, and user interfaces run in software, whereas the FPGA is used to collect sensor data, extract features, and to calculate the nearest neighbor on the extracted information. The system runs at 160\,MHz consumes 4.96\,mW on the CPU part and 153.97\,mW on the FPGA side.
\textcolor{black}{While their power consumption is within range of IoT applications, these FPGAs are limited in performance and thus not suitable for computationally intensive applications.}
To enrich the functionalities of deeply embedded \acp{SoC}, FPGA vendors started to develop and commercialize FPGA IPs that can be integrated into \acp{SoC}, presented in the following section.

\subsection{eFPGAs}

eFPGAs are FPGA IP cores specifically meant to be integrated into \acp{SoC} to extend them with programmable logic. Unlike the FPGAs presented in the previous section, eFPGAs are not meant to be used standalone, but are designed with the goal of enhancing the capabilities of the \acp{SoC}. Vendors provide tools to allow eFPGAs to be customized to the \acp{SoC} and properties like the number of arrays, with a given number of LUTs, DSP blocks, flip-flops, \io{} pins, etc. can be configured. eFPGAs can be provided as soft-IP \cite{Renzini, menta}, described in RTL and synthesized with the rest of the system, or hard-IP \cite{Whatmough, Borgatti, Lodi} as hard-macros with pre-determined physical layout, featuring a different trade-off between performance and cost.
Although soft eFPGA macros are easily portable from different technology nodes as they are made by standard cells, hard-macro eFPGAs, which are usually custom-designed at layout level, feature significantly better \ac{PPA} figures.

For example, in Renzini et al. \cite{Renzini}, a soft-IP is complementing a \ac{MCU} for power control applications is implemented using a 90\,nm Bipolar CMOS DMOS (BCD) technology. This eFPGA is relatively small (only 96 4-input LUTs and 192 flip-flops) and connected exclusively to the \io{} subsystem to implement low-latency and flexible control tasks such as Pulse Width Modulation (PWM). Several companies are providing hard-IP blocks, as Achronix \cite{achronix}, which provides 7nm FinFET eFPGAs, Flex-Logix \cite{flexlogix}, which provides from 12\,nm to 180\,nm eFPGAs macros, QuickLogic Corporation \cite{quicklogic}, which provides from 22\,nm to 65\,nm core IPs, and Menta \cite{menta}, which provides IPs from 10\,nm to 90\,nm.
Several heterogeneous reconfigurable \acp{SoC} have been presented in the last years, ranging from high-performance systems to low-power embedded systems. Whatmough et al. presented a 25\,mm\textsuperscript{2} \ac{SoC} implemented in 16\,nm FinFET technology featuring two ARM A53 cores, a quad-core datapath accelerator, 4\,MBytes on-chip SRAM, and a 2 $\times$ 2 FlexLogic eFPGA macro featuring hardwired DSP slices \cite{Whatmough}. The proposed SoC can achieve up to 28.9$\times$ better energy-efficiency when DSP and crypto algorithms are executed on the eFPGA rather than the ARM cores.

In the embedded domain, several solutions have been proposed in different technology nodes. Borgatti et al. \cite{Borgatti} implemented a 180\,nm 2\,0mm\textsuperscript{2} \ac{SoC}, where eFPGA is integrated with the CPU pipeline to implement a reconfigurable Application Specific Instruction Processor (ASIP) SoC, with the eFPGA implementing custom instructions. In addition, the eFPGA is connected to the system bus and \io{} pads. The system reports up to 10$\times$ performance gain using instruction extensions to accelerate face-recognition algorithms and 2$\times$ for \io{} intensive tasks when dealing with camera peripherals with pre-processing. Lodi et al. \cite{Lodi} implemented a 42\,mm\textsuperscript{2} \ac{SoC} in 130\,nm, where the CPU pipeline is directly connected with the eFPGA to implement custom instructions, whereas a second eFPGA is connected to the system bus and \io{} pads. The system reports up to 15$\times$ performance gain and 89\% energy saving by exploiting the eFPGAs to accelerate a set of data processing algorithms. However, as a consequence of using a mature technology node, the eFPGAs (\textasciitilde{}15\,kGE) presented in the proposed SoCs feature limited capabilities and performance.

To boost signal processing workloads, both hard and soft eFPGAs can have \ac{DSP}-blocks included in the IP itself, or they can have pins dedicated to communicating with external blocks, featuring, once again, a different trade-off between time to market for DSP-blocks customization at design time. The first ones can be used by eFPGA synthesis tools to map user-designs in \ac{DSP}-blocks implicitly, whereas in the second case, the user explicitly designs logic in the eFPGA to interact with the external blocks. All the works featuring \ac{DSP}-blocks so far belong to the first category, whereas the proposed work has MAC-blocks external to the IP macro.


In this work, we propose an SoC featuring an advanced microcontroller augmented by an embedded eFPGA for IoT applications in 22\,nm process technology. Differently from what has been proposed in Whatmough et al. \cite{Whatmough}, we target a much lower power budget.
The proposed SoC utilizes the eFPGA to enhance the \io{} capabilities of the \ac{SoC}, by performing I/O pre-processing tasks as well as being used as a tightly coupled accelerator. The proposed solution provides 3.4$\times$ better performance and 2.9$\times$ better efficiency than state-of-the-art heterogeneous reconfigurable SoCs. One key feature of the SoC is the unique capability to exploit reverse body biasing enabled by FD-SOI technology to implement a 20.5\,\textmu{W} state-retentive deep-sleep mode for the eFPGA. This point is further discussed in Section~\ref{sec:physicaldesign}.

\begin{figure}
    \centering
\centerline{\includegraphics[width=\linewidth]{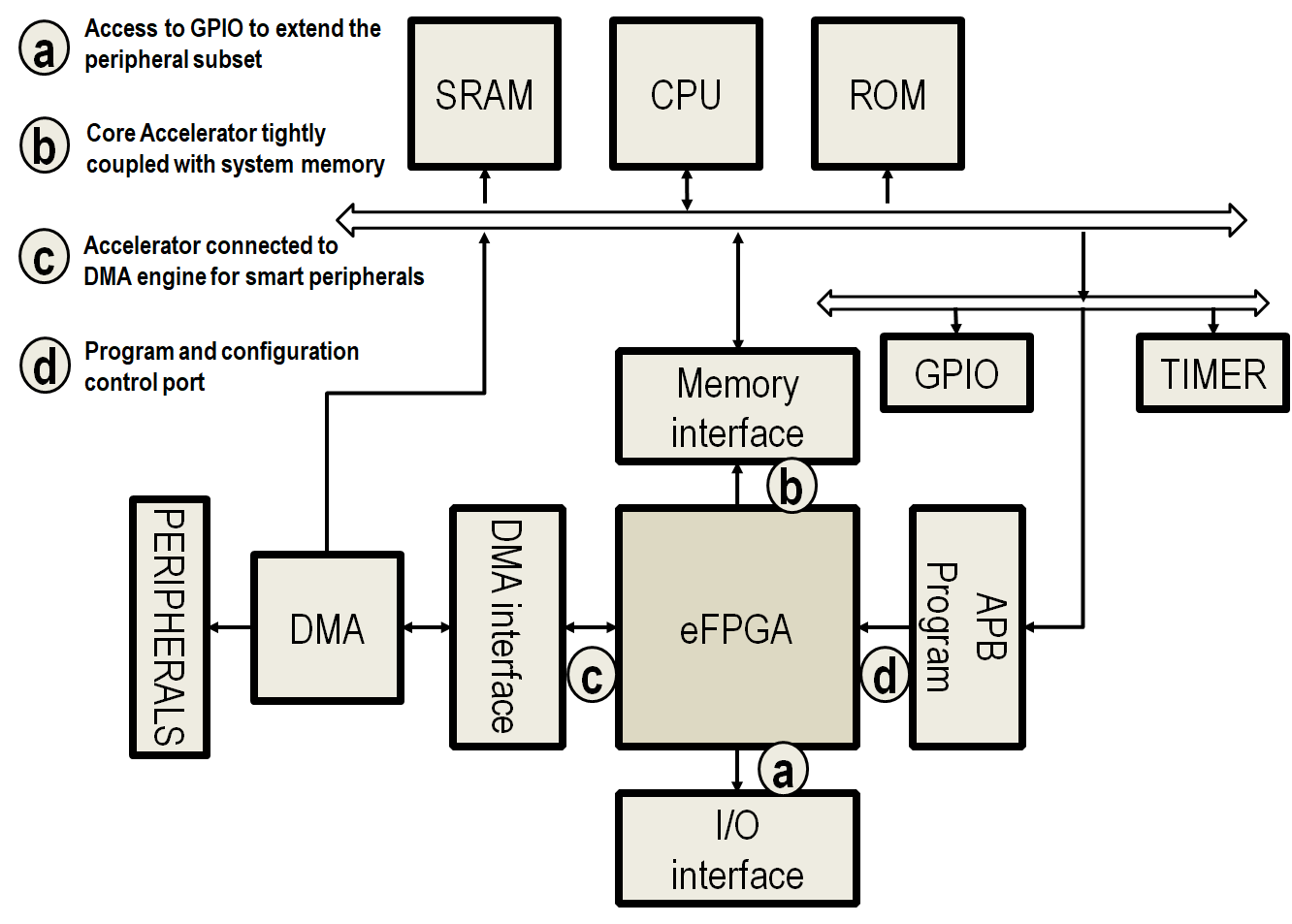}}
    \caption{\ac{MCU}-eFPGA SoC architecture. eFPGA connections towards the \ac{MCU} and to the external peripherals are highlighted.}
    \label{fig:architecture}
    \vspace{-4mm}
\end{figure}

\section{Arnold Architecture}
\label{sec:architecture}


\begin{figure*}
    \centering
\centerline{\includegraphics[width=1.00\textwidth]{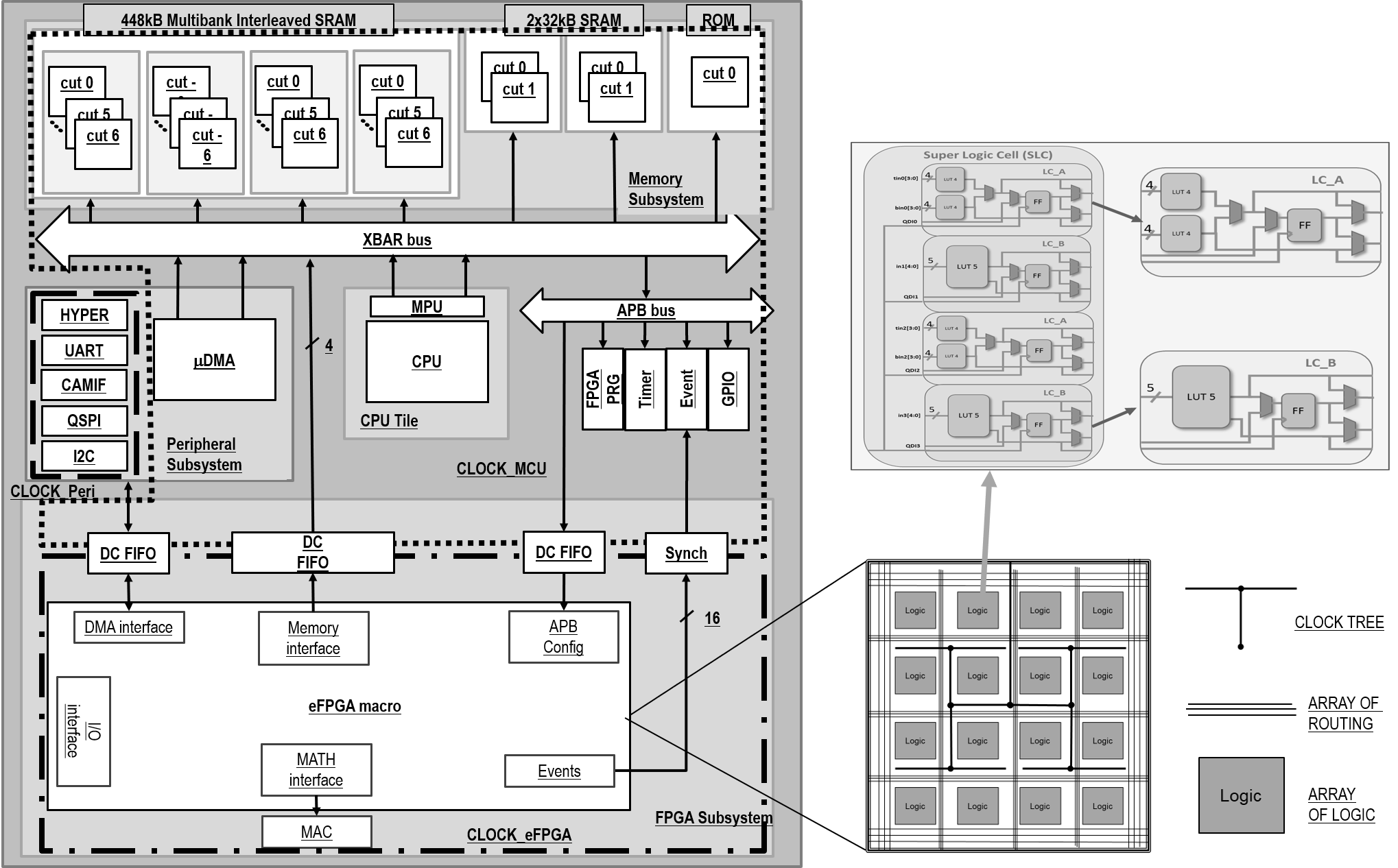}}
    \caption{Detailed block diagram of the proposed design. The eFPGA (bottom) connected with the \ac{MCU} and its private MAC units in a clock domain (CLOCK eFPGA). Peripherals (center – left) are directly connected to the \textmu DMA in the Peripheral subsystem and operate on the CLOCK Peri clock domain. The rest of the system works in the CLOCK \ac{MCU} domain. The CPU runs the SW and orchestrates the whole system.}
    \label{fig:blockdiagram}
    \vspace{-4mm}
\end{figure*}


The proposed system is built around an in-order RISC-V core\footnote{The OpenHW Group CV32E40P is freely downloadable at https://github.com/openhwgroup/ under the SolderPad license.} based on \cite{gautschi2017near}, optimized for signal processing, featuring a 4-stage pipeline, and achieving 3.19\,Coremark/MHz and up to 2.4\, eight-bit GMAC/s (at 600\,MHz). The core implements the RISC-V 32\,bit integer (I), multiplication and division (M), single-precision floating-point (F), and compressed (C) \ac{ISA} extensions (RV32IMFC) \cite{Waterman14theriscv}. In addition, the core has been extended with custom instructions to speed up data processing applications such as zero-overhead hardware loops, automatic increment load/store instructions, bit manipulations, and packed-single-instruction-multiple-data (pSIMD) operations between vectors of 4 bytes or 2 half-words at a time.

To protect sensitive parts of the system from corrupted user applications, we extended the CPU with a RISC-V compliant \ac{PMP} unit that can control read, write, and execute permissions on regions of the physical memory. The implemented RISC-V \ac{PMP} supports all address matching schemes as: naturally aligned power of 2 regions \textit{NAPOT} (including 4 bytes alignment \textit{NA4}); and the top boundary of an arbitrary range \textit{TOR}. The \ac{PMP} occupies only \risckyPMPovh{} of the total CPU area due to the extra registers and comparators needed to implement the specifications and provides much-needed security features for user-applications in the IoT domain. In the proposed SoC, the CPU is responsible for executing the runtime to manage the system and to execute user applications to process data or to control external peripherals, as well as to configure and control the eFPGA itself.

\subsection{Memory Subsystem}

The memory system, composed of 512\,kB of \ac{SRAM}, is shared among the CPU (instruction and data), the \io{} DMA ($\mu$DMA) (RX and TX), the JTAG, and the eFPGA masters. The memories are slaves of the system bus, which is based on a single-cycle latency logarithmic interconnect \cite{LogInt} (XBAR bus in \figref{fig:blockdiagram}). In case two or more masters request to access the same slave, a round-robin arbiter selects the master that first communicates with the slave to solve the conflict. The shared memory consists of four word-level interleaved memory banks, each with 112\,kB each, and two memory banks of 32\,kB featuring a non-interleaved address scheme. Every memory bank is a composition of single-port 4096 by 32\,bit words (16\,kB) memory cuts optimized for density and power. The size chosen for the memory cuts allows to place them comfortably during the physical implementation as described below, and concurrently to meet the frequency target.

The chosen interleaving scheme for the four 112\,kB (448\,kB) memory portion approximates a multi-port memory access, and it increases the bandwidth up to 4$\times$ when multiple masters are loading or storing data sequentially, which is the typical case for most \ac{DSP} applications. When low-latency single-cycle accesses with no contention are needed, the two private banks can be used, which offer a bandwidth of 19.2\,Gbps each. In the proposed \ac{MCU}, they are used to store private CPU data such as the stack and instruction binary. In this way, the interleaved part can be used by the other masters with no conflicts. This solution avoids the use of power and area hungry multi-port memory cuts, still providing low-latency access to memory, increasing the total energy efficiency. A \ac{ROM} has also been implemented to store the boot instructions responsible for setting the system upon reset.

\subsection{\io{} subsystem}

The \io{} subsystem is composed of a broad set of peripherals that include JTAG, HyperRam, UART, Camera Interface, quad-SPI, and I2C, which communicate with the shared memory system through an autonomous \textmu DMA based on \cite{pullini2017mudma}. The \textmu DMA is a smart-engine that allows peripherals to control transfers to/from memory without the need for the CPU continuous control. The HyperRam peripheral is particularly interesting as it allows to access off-chip memory with a bandwidth of 800\,Mbps, extending the \ac{MCU} with larger memory capacity, useful for holding several eFPGA bitstreams.

The \textmu DMA has two ports towards the main memory, one to transmit and one to receive data from peripherals. At 600\,MHz, the \textmu DMA has an aggregated bandwidth equal to 38.4\,Mbps. Except for the JTAG, which is directly connected to a master port of the system bus, the other peripherals are controlled by the \textmu DMA core, which handles memory requests in a time-multiplexed fashion. The \textmu DMA control registers are used to select the active peripheral, the peripheral clock frequency, number of transfers, etc. Other peripherals, such as SoC control registers, timers, GPIOs, and event units are also included in the proposed \ac{MCU} and accessible through the \ac{APB} bus.

\subsection{Clock subsystem}

\textit{Arnold} includes three \acp{FLL} that take as input an external 32\,kHz reference clock and provide internal clocks up to \maxFLL{}. One \ac{FLL} each is used to provide the clock to the eFPGA, the peripheral subsystem and the remaining modules as CPU, memories, busses, etc. The eFPGA has access to six clock sources: four from external GPIOs; one from the eFPGA \ac{FLL} block; and one from an integer frequency divider from the same \ac{FLL}.

\subsection{eFPGA subsystem}

The eFPGA is tightly coupled to the system to minimize the overhead of communications with the CPU. It has 3712 pins to be used to connect the IP with the rest of the \ac{SoC}. In this work, we designed a novel, highly flexible 4-mode  \ac{SoC} interface to:
 \begin{enumerate}[label=(\alph*)]
   \item an \io{} interface with direct connections toward the pad frame of the system, enabling the implementation of custom off-chip interfaces;
   \item a memory interface suitable for shared-memory accelerators implemented on the FPGA logic and tightly coupled with the CPU;
   \item an \io{} DMA interface suitable for implementing \io{} filtering functions for data streamed into the system from the standard \io{};
   \item an \ac{APB} configuration and control interface suitable for controlling the programmable logic.
 \end{enumerate}

The \io{} interface is made of 41 sets of three signals (input, output, direction) from the eFPGA to the GPIOs. This interface is used for custom \io{} protocols, which are challenging to implement efficiently in SW due to latency constraints.
Each \io{} pad can be either used by a peripheral (quad-SPI, Camera Interface, etc.), or by software (Core GPIO), or by the eFPGA. Multiplexers controlled by SoC registers drive the functionality mode of each pad.

The memory interface implements the protocol presented in \cite{LogInt}. The proposed \ac{SoC} has four interfaces connected as master ports in the bus, providing up to 128\,bit memory operations (load or store) per transaction. Access to the on-chip \ac{SRAM} is provided through four 32\,bit 4\,words dual-clock FIFOs to allow the \ac{MCU} and the eFPGA subsystem to operate at independent frequencies. This is a crucial feature since the eFPGA usually runs at a lower frequency than the rest of the SoC and its frequency depends on the user design. For security reasons, the eFPGA memory interface has only access to \ac{SRAM} banks and not to \ac{APB} peripherals and boot \ac{ROM}.

The \io{} DMA interface is composed of one receive (RX), and one transmit (TX) bus featuring a ready/valid handshaking, plus one 32\,bit configuration bus as described in \cite{pullini2017mudma}. The configuration bus allows controlling the peripherals mapped into the eFPGA with external registers which can avoid the use of the \ac{APB} interface described below, and thus save resources. In addition, this interface can be used to stream data through the \textmu DMA without using eFPGA resources for the address generation logic as it would with the memory interface. In this case, the \textmu DMA transfers data from the eFPGA to memory (and vice versa) linearly. Communication between the \textmu DMA and the eFPGA happens using two 32\,bit 4\,words dual-clock FIFOs.

Designs mapped into the eFPGA (as accelerators or peripherals) can be controlled by registers through the \ac{APB} configuration and control interface. Such an interface is made of a 7\,bit address, 32\,bit data read, and data write, write-enable, ready, peripheral select and enable signals (75 pins). One 32\,bit 4\,words dual-clock FIFO is used for communications between the \ac{MCU} and the eFPGA.

In addition to the four interfaces mentioned above, the eFPGA can generate sixteen events to interact asynchronously with the CPU, avoiding inefficient polling operations and saving power. In fact, the eFPGA event pins are connected to dual-clock event-propagators that notify the events to the CPU as dedicated interrupts requests. The interrupt service routines are user-defined, and they can be used to handle the eFPGA requests, for example, starting a new \io{} transaction, or programming the new acquired data pointers to start processing them in case of accelerator design.

To improve computational arithmetic density, two synthesizable parallel-vectorial \ac{MAC} accelerators are connected to the eFPGA to compute four 8\,bit, two 16\,bit, or one 32\,bit \ac{MAC} operations for each unit. The two MAC blocks are connected via 310 pins each, which control the MAC blocks, whether data comes from the eFPGA or the MAC buffers, the input and output data, and the vector mode (8, 16, or 32).

The CPU programs the eFPGA through another \ac{APB} interface. Such master interface is connected to the eFPGA \ac{FCB}, which is responsible for controlling the eFPGA, managing the power procedures, and report the actual status of the eFPGA. The eFPGA binary is 225.5\,kB, small enough to be contained in the on-chip \ac{SRAM}. To program the macro, the CPU reads the binary from an external memory to the on-chip memory, then the CPU reads the binary array and writes its content to the \ac{APB} \ac{FCB} via non-critical load and store instructions.

The eFPGA fabric is organized in four quadrants with dynamic reconfiguration capabilities, each one composed of an array of 16x16 Super Logic Cells (SLCs). Each SLC has four logic cells that are organized in two sub-logic clusters: two instances of logic cell A (LCA) and two instances of logic cell B (LCB), as shown in \figref{fig:blockdiagram}. Both LCA and LCB also include one register and multiple multiplexers that enable the logic cell to perform different functions (e.g., combinatorial, sequential, or both). If a logic cluster or a highway network within the SLC is not used, it is powered off to save static power. A shared register clock, set, and reset signals for all four logic cells helps reduce routing congestion. If the logic cluster or highway network within the SLC is not used, it is powered off to save static power.

\section{eFPGA Software and Tools}
\label{sec:software}

To use the eFPGA in the \textit{Arnold} SoC, the user writes HDL code (VHDL, Verilog or SystemVerilog) and synthesizes it with Mentor Graphics Corporation \textsuperscript{\textcopyright} Precision RTL Synthesis OEM Quicklogic tool. The synthesized design is then placed and routed with the QuickLogic Aurora Software Tool Suite (Aurora). The user must map each of the soft-module interface pins to the corresponding pin of the eFPGA hard-macro. For example, the user may define the memory interface request signal as ``MemREQ\_output'', in the Aurora tool, the user may specify that the signal is connected to the 3rd memory interface of the eFPGA specifying that ``MemREQ\_output'' is connected to ``tcdm\_req\_p3\_o'' pin.
The eFPGA pin has been assigned to its interface functionality at SoC design time to optimize the place and route phase.

Once the constraints and the pin mapping have been defined, Aurora performs logic optimization on the synthesized design, places, and routes it. It also generates static timing analysis and the bitstream containing the binary of the user-design. The binary is then loaded into the main memory by the CPU. The CPU stores each binary word into the bitstream registers. Once the eFPGA has been programmed, the CPU can control the design with user-defined registers mapped into the eFPGA \ac{APB} interface described above to start the design, to check the status, etc. \acp{API} have been developed to provide C procedures for the user. In particular, functions to RESET the eFPGA, to load the bitstream, and to wait for the end of the eFPGA computation (\textit{wait\_fpga\_eoc}) have been implemented for fast integration into the user application. The \textit{wait\_fpga\_eoc} routine leverages the ``wait for interrupt'' (WFI) RISC-V instruction to clock-gate the CPU to save dynamic power.

\section{Arnold Physical Design}
\label{sec:physicaldesign}

The proposed \ac{SoC} fabricated in \ac{GF22FDX} 10 Metal technology occupies 3$\times$3 mm\textsuperscript{2}. \ac{FDSOI} technology has been chosen as it provides performance and power knobs through body-biasing, and it is highly energy-efficient over a wide  Vdd range \cite{zaruba2019floating} as confirmed by our results discussed in the Subsection~\ref{subsec:perf}.
The synthesis tool used for this project is Synopsys \textsuperscript{\textcopyright} Design Compiler 2017.09, whereas the place and route tool used is Cadence \textsuperscript{\textcopyright} Innovus 18.11. The design has been closed at 430\,MHz for the \ac{MCU} side and for up to 100\,MHz for the eFPGA soft-designs. Worst-case conditions at 0.72\,V for setup constraints, and best-case conditions at 0.88\,V for hold constraints between -40$^{\circ}$C and 125$^{\circ}$C have been used to guarantee performance across the process, voltage, and temperature variations.

\begin{figure}
    \centering
\centerline{\includegraphics[width=\columnwidth]{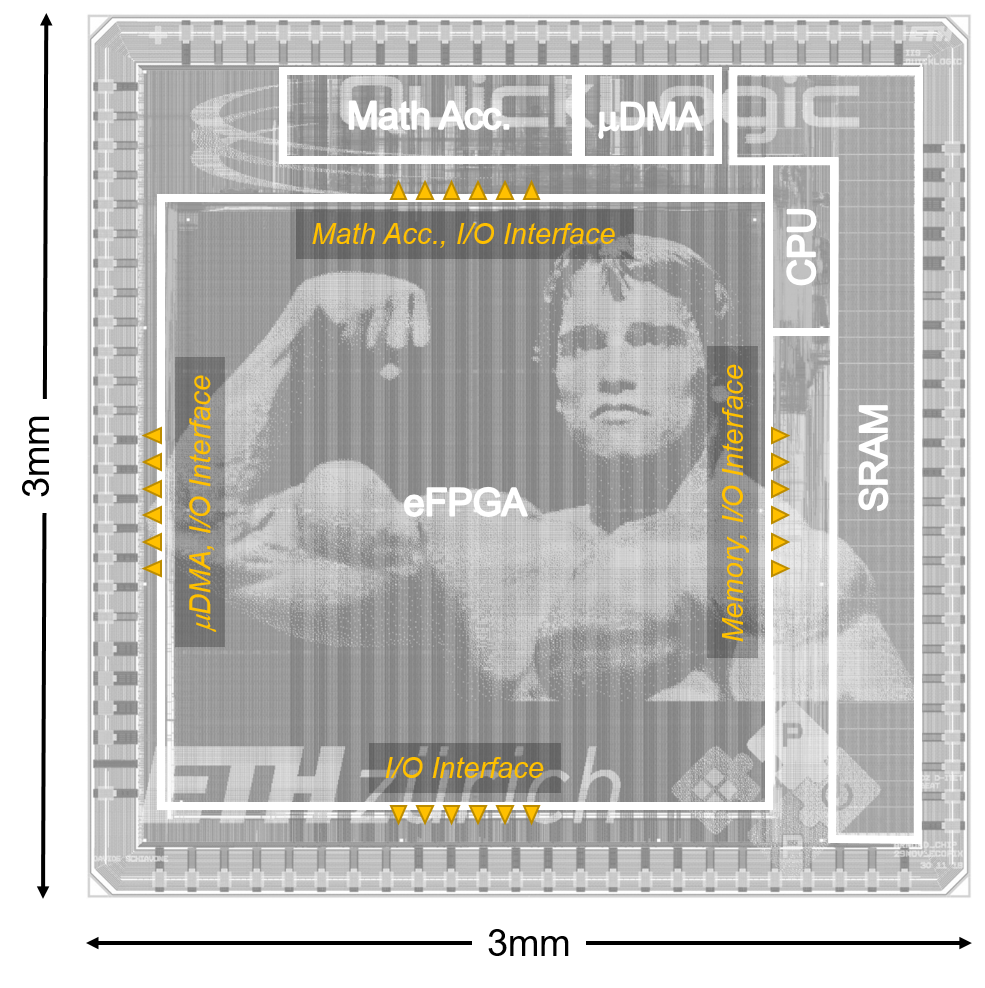}}
    \caption{ Die photo of the proposed design with the main components and eFPGA pins highlighted.}
    \label{fig:diephoto}
    \vspace{-4mm}
\end{figure}

The die picture and floorplan of the chip are shown in \figref{fig:diephoto}. The eFPGA macro is 2$\times$2 mm\textsuperscript{2}, and it has been placed in the bottom left of the design. The memory cuts have been placed to the right of the eFPGA. The eFPGA memory interface pins have been assigned to the right part of the eFPGA to minimize routing efforts and to minimize the congestion issue as the path towards the memory is the most critical. The core has also been automatically placed close to the memory to minimize timing penalties. The eFPGA pins for the MAC blocks accelerators have been placed to the top part, where the local math accelerator \ac{SRAM} buffers have been placed. On the left part of the eFPGA, the pins towards the \textmu DMA, the user \ac{APB} interface, and the 16 events pins have been assigned. GPIOs pins are spread along the four sides of the eFPGA. The six clock pins of the eFPGA are located three on the top and three on the bottom side. The three \ac{FLL}s have been placed on the top part of the chip, whereas the standard cells have been automatically placed by the place and route tool.

The effective area occupied by the chip is \effectiveArea{}, of which the eFPGA macro occupies 78\% (4mm\textsuperscript{2}) and the \ac{MCU} 22\% (\socareanoefpga{}).
The main memory occupies 14.46\% of the system area, whereas the \io{} subsystem and the CPU take only 0.43\% and 0.54\%, respectively. The eFPGA subsystem components occupy 1.26\% of \ac{MCU} area. The eFPGA subsystem is a set of modules that interact directly with the eFPGA macro, dual-clock FIFOs, the \ac{FCB}, the MAC accelerators (including memory buffers), and clock multiplexing logic. Table \ref{tab:area_distr} shows the area distribution of the chip. 

\begin{table}
\centering
   \begin{threeparttable}
        \caption{Area distribution of the main components of Arnold.}

       \begin{tabular}{@{}lrr@{}}
            \toprule
                \textbf{Module} & \textbf{Area [\textmu m\textsuperscript{2}]} & \textbf{Percentage} \\\toprule

                CPU        &   27'186  &  0.54\% \\
               Main Memory        &  734'232  & 14.46\% \\
               I/O DMA            &   21'755  &  0.43\%  \\
               eFPGA subsystem &   63'946  &  1.26\%  \\
               PAD Frame       &  229'519  &  4.52\%  \\
                eFPGA Macro        & 4'000'000  & 78.79\%   \\\bottomrule
        \end{tabular}
    \label{tab:area_distr}
    \end{threeparttable}
\vspace{-4mm}
\end{table}

The \ac{MCU} and the eFPGA operate at the same supply voltage, but the eFPGA can be switched off from external power managers. The range of operation is between 0.5\,V to 0.8\,V. To reduce the leakage power while preserving the eFPGA configuration during state-retentive deep sleep states, RBB is applied from an external generator to minimize on-chip implementations overheads. On the other hand, forward body-bias (FBB) is applied to the CPU, memory, and the rest of the logic to increase performance \cite{schiavone2018quentin, di2019pushing}.

\subsection{Performance and Energy Efficiency}
\label{subsec:perf}

\begin{figure*}
    \centering
\centerline{\includegraphics[width=\linewidth, scale=0.1]{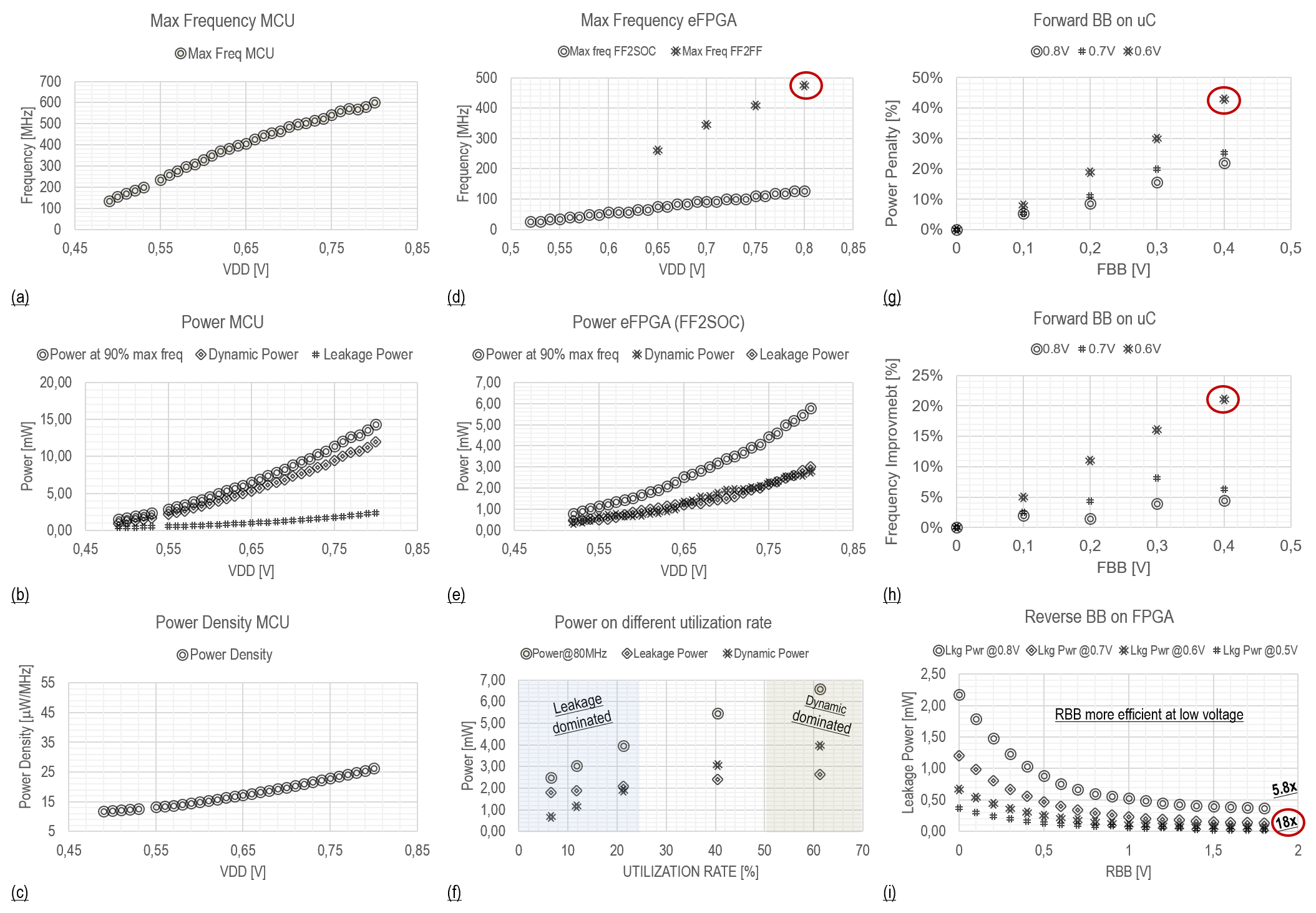}}
    \caption{Frequency (a), power consumption (b), and energy-efficiency (c) with respect to the supply voltage of the MCU part of the proposed design. In the center, frequency (d) and power of the eFPGA macro with respect to the supply voltage (e) and power with respect to the utilization rate (f). The effect of the FBB on power (g) and frequency (h) on the MCU. The effect of RBB on the eFPGA leakage power during state-retentive
deep-sleep mode (i).}
    \label{fig:meausurements}
    \vspace{-4mm}
\end{figure*}

In this subsection, measured results at room temperature from the implemented chip are reported and discussed. Performance and power results have been measured using an Advantest SoC V93000  ASIC tester. \figref{fig:meausurements} (left) shows the maximum frequency (a), power consumption (b), and power density (c) of the \ac{MCU} during the execution of a matrix multiplication at different supply voltages. Measured results at ambient temperature show a maximum frequency of 135\,MHz and power consumption 11.88\,\textmu W/MHz at 0.49\,V, up to a maximum of 600\,MHz at the nominal 0.8\,V while consuming 26.18\,\textmu W/MHz. The maximum frequency at 0.49\,V is comparable with commercial single-core MCUs performance while achieving very low power consumption thanks to voltage scaling. When high performance is needed, 600\,MOPS can be achieved at a maximum power consumption of 16\,mW. The leakage power of the whole \ac{MCU} ranges from 0.53\,mW (33\%) to 2.39\,mW (15\%) at 0.49\,V and 0.8\,V respectively. \figref{fig:meausurements}(g) shows the effect of the FBB on the MCU power consumption, and \figref{fig:meausurements}(h) on the frequency. The MCU can run up to 20\% faster at 0.6\,V at the price of 43\% higher power consumption, whereas the effect of FBB is smaller when applied at 0.8\,V (only 5\% faster) for a maximum frequency of 630\,MHz. The effect of the magnified impact of body biasing at low voltage is a well-known effect seen in near-threshold FD-SOI chips \cite{ROSSI2016170}.

\figref{fig:meausurements} (center) \textcolor{black}{shows the eFPGA measured results.
\figref{fig:meausurements}(d) shows the maximum frequency of two different designs: \textit{FF2SOC} is an eight-way parallel 32\,bit accumulator that reads values from the SoC memory and accumulates them in eight different registers. The signature can be read with the APB interface; \textit{FF2FF} is a nine bit counter that divides the eFPGA clock by 512 and drives a GPIO with the divided clock.
The designs are different as the \textit{FF2SOC} communicates with synchronous elements in the SoC (dual-clock FIFOs), and thus its maximum frequency is bounded by the internal delays of the eFPGA and the logic outside its boundary, whereas \textit{FF2FF} has been designed to measure only the flip-flop to flip-flop delay, without taking into account the propagation and setup timing of the eFPGA and the external logic at its boundary. The output of the Q-pin of the MSB flip-flop of the nine bit counter is directly connected to the GPIO, and the frequency is measured with an oscilloscope. From measurements we determined a maximum frequency of 475\,MHz at 0.8\,V and 260\,MHz at 0.65\,V.
\textit{FF2SOC} occupies 15\% of the internal eFPGA resources and it can run from 26.38\,MHz, consuming 34.34\,\textmu W/MHz at 0.52\,V, to 126.88\,MHz at 0.8\,V consuming 47.98\,\textmu W/MHz (\figref{fig:meausurements}(e))}.

The eFPGA \textit{FF2SOC} leakage power is 0.38\,mW at 0.5\,V, up to 2.18\,mW at 0.8\,V.  The power has been measured separately from the rest of the system as the power grid stripes of the eFPGA are different from the MCU ones. The power overhead added by the eFPGA is affordable in the IoT domain, making the integration of such programmable arrays a viable option for the next generation of edge-computing nodes. The eFPGA leakage power consumption is reduced via state-retentive deep sleep states applying RBB, resulting in a minimum leakage power of 20.5\,\textmu W at 0.5\,V and 374.2\,\textmu W at 0.8\,V and 1.8\,V reverse body-bias as shown in \figref{fig:meausurements}(i), i.e., a 5.8$\times$( at 0.8\,V) to 18$\times$ (at 0.5\,V) reduction can be achieved thanks to RBB. This result makes the eFPGA power consumption significantly reduced when not used, minimizing the integration cost and overhead. \figref{fig:meausurements}(f) shows how the power consumption changes with respect to the utilization rate. A design with a parametrizable number of adders has been implemented in the eFPGA to measure the power consumption with respect to the utilization rate. When running at 80\,MHz, 0.75\,V, results show an energy-efficiency of 0.40\,\textmu W/MHz/SLC, being leakage dominated when <20\% of resources are utilized. The best energy-efficient point of the whole system is 46.83\,\textmu W/MHz (eFPGA consumes 28\% of total power) achieved in near-threshold at 0.52\,V, when the core and the eFPGA are running at 183.6\,MHz and 26.38\,MHz respectively. This result has been measured when the eight parallel 32\,bit accumulators are mapped on the eFPGA.

\begin{table*}
    \begin{threeparttable}
    \caption{Performance comparison with state-of-the-art MCU and eFPGA systems.}
    \begin{tabularx}{\linewidth}{@{}l | ccccccc @{}}
        \toprule
        &
        \textbf{Borgatti}  &
        \textbf{Lodi}  &
        \textbf{Renzini}  &
        \textbf{Fournaris}  &
        \textbf{Whatmough}  &
        \textbf{Bol} &
        \textbf{This} \\
        &
        \cite{Borgatti} &
        \cite{Lodi} &
        \cite{Renzini} &
        \cite{Fournaris} &
        \cite{Whatmough} &
         \cite{Bol19} &
        \textbf{Work} \\

        \midrule

        Technology [nm] &
        180      &
        130      &
        90       &
        65      &
        16      &
        28 &
        22 \\

        I\$/D\$/SRAM [kB] &
        8/8/48  &
        8/8/256 &
        -/-/32  &
        8/-/656 &
        2K\tnote{$^1$}/-/4K &
        -/-/64  &
        -/-/512\\

        Voltage Range [V] &
        1.8  &
        1.2  &
        1.2  &
        1.2  &
        0.5 - 1.0  &
        0.4 - 0.8 &
        0.5 - 0.8\\

        FPGA IP macro &
        Hard &
        Hard &
        Soft &
        Hard &
        Hard &
        - &
        Hard \\

        FPGA Area [mm\textsuperscript{2}]  &
        8.2   &
        6.0 &
        0.347 &
        -     &
        1.0 &
        - &
        4.0 \\

        FPGA \#LUT   &
        15\,kGE  &
        15\,kGE  &
        96 5/4:2 &
        12084 4:1 \tnote{$^2$}  &
        8800 6:2 \tnote{$^3$} &
        - &
        6018 4:1 \\

        FPGA \#FF   &
        -  &
        -  &
        192 &
        12084 \tnote{$^2$} &
        22656         \tnote{$^4$}  &
        -  &
        4096 \\

       FPGA \#DSP   &
        -  &
        -  &
        -  &
        22 \tnote{$^5$} &
        80 MACs \tnote{$^6$} &
        - &
        2 vecMACs  \\

    Access Mode to  &
    GPIOs   &
    GPIOs       &
    s mmap  &
    GPIOs &
    m/s mmap &
    - &
    GPIOs \\

     SoC &
    m/s mmap   &
    s mmap       &
      &
    m/s mmap &
     &
    - &
    m/s mmap \\

      &
    RX DMA   &
    TX/RX DMA       &
     &
    TX/RX DMA &
    &
    - &
    TX/RX DMA \\

     FPGA Lkg Power \tnote{*} &
    -  &
    -  &
    -  &
    -  &
    12000 \tnote{$^7$} &
    - &
    20.5 - 2178 \\

     FPGA Max Freq.\tnote{**}   &
    175  &
    166  &
    50   &
    160  &
    734   &
    - &
    475 \\

     FPGA Power &
    -  &
    -  &
    34.72@1.2V \tnote{$^8$} &
    962@1.2V \tnote{$^9$}&
    - &
    - &
    31.98@0.6V \tnote{$^{10}$} \\

      Density \tnote{***} &
    &
     &
    &
    &
     &
     &
     \\

     MCU Lkg Power \tnote{*} &
    -  &
    -  &
    -  &
    7000\tnote{$^{11}$}  &
    - &
    1 - 30
    & 532 - 2386 \\

     MCU Max Freq. \tnote{**} &
    175  &
    166  &
    50   &
    166 &
    - &
    80 &
    600 \\

     MCU Power     &
    -  &
    -  &
    101.22@1.2V\tnote{$^8$} &
    31@1.2V\tnote{$^{12}$}  &
    -  &
    3@0.4V, &
    11.88@0.49V,\\

    Density \tnote{***}   &
    &
    &
    &
    &
    &
    48MHz &
    135MHz \\

     MCU+FPGA    &
    -  &
    1807.23@1.8V &
    135.94@1.2V \tnote{$^8$} &
    993@1.2V \tnote{$^9$,$^{12}$} &
    -  &
    - &
    46.83@0.52V \tnote{$^{13}$} \\

   Power Density \tnote{***}   &
    &
    &
    &
    &
    &
    &
     \\

    \bottomrule

    \end{tabularx}
    \begin{tablenotes}
    \scriptsize{}
    \item [*] Power numbers are in \textmu W
    \item [**] Frequency numbers are in MHz
    \item [***] Power density numbers are in \textmu W/MHz
    \item [1] Two 64\,kB of L1 cache shared between Instructions and Data for each core, plus 2\,MBytes of L2 cache. \\
    \item [2] SmartFusion2 M2S010S data available in the product brief.\\
    \item [3] 2520$\times$2 LUTs for the two logic tile and 1088$\times$2 for the two DSP tiles \cite{EFLX16}. \\
    \item [4] 6304$\times$2 flip-flops for  two logic tile. 5024$\times$2 for the two DSP tiles \cite{EFLX16}.\\
    \item [5] Signed multiplication, dot product, and built-in addition, subtraction, and accumulation units. \\
    \item [6] 40$\times$2 MACs for the two DSP tile\cite{EFLX16}. \\
    \item [7] 3\,mW reported in the datasheet \cite{EFLX16}. Assuming it is for a 1$\times$1 tile,  \cite{Whatmough} uses a 2$\times$2 tile, thus 12\,mW have been reported in the Table.  \\
    \item [8] Average measurements. \\
    \item [9] Estimated from \cite{Fournaris}. It assumes the FPGA runs at 160\,MHz.
    \item [10] When \textit{FF2SOC} design is synthesized on the eFPGA
    \item [11] Includes FPGA leakage power as well. \\
    \item [12] Number taken from \cite{Fournaris}. The authors use the ARM Cortex M3 power consumption from the datasheet reported in 90\,nm LP.\\
    \item [13] When \textit{FF2SOC} design is synthesized and running on the eFPGA and the MCU is computing a matrix multiplication at the same time
    \end{tablenotes}
    \label{tab:soa_table}
    \end{threeparttable}
\end{table*}

\section{Use Cases}
\label{sec:usecase}

To demonstrate the flexibility and efficiency of our heterogeneous reconfigurable SoC, three different use cases have been implemented, highlighting the versatility offered by embedded programmable logic.

\begin{table}
    \begin{threeparttable}
        \caption{Resource utilization, power consumption and overall energy savings for implementing different use-cases on the eFPGA.}

        \begin{tabular}{@{}lrrrrr@{}}
            \toprule
                \textbf{Use Case} &
                \textbf{GPIO} &
                \textbf{FF}  &
                \textbf{LUT}  &
                \textbf{Power}  &
                \textbf{Energy}\\
                & & & & \textbf{[mW]} & \textbf{Saving [$\times$]}
                                \\\toprule

                Custom \io{}  & 36  & 205  & 289 & 6.0 & 2.5\\
                BNN         & 0 & 854  & 1229  & 12.5 & 2.2\\
                CRC         & 0 & 20  & 47  & 7.5 & 42.2 \\

                \bottomrule
        \end{tabular}
    \label{tab:use_case}
    \end{threeparttable}
    \vspace{-4mm}
\end{table}

\subsection{\io{} subsystem accelerator}

In the context of applications for bio-signal processing, it is common to extract features in the frequency domain to classify activities sensed from skeletal muscles or the brain \cite{kartsch2018sensor}. Wavelet or Fourier transforms are used to convert the signal from the time to the frequency domain, then features like the spectral power, are extracted and used by a pattern recognition algorithm. For this reason, a peripheral that extracts relevant information of the signal acquired from the sensors has been developed and mapped to the eFPGA to alleviate the pre-processing part of the CPU, which then classifies the activity starting from the extracted features. The peripheral accelerator mapped on the eFPGA consists of an SPI module extended with computational capabilities to calculate the Haar Discrete Wavelet (HDWT), which is an attractive algorithm to implement in an eFPGA as it does not require multipliers \cite{elfouly2008comparison}.

The accelerator is configured to acquire $N$ samples of 16\,bit of raw data coming from ADCs, and to store the Approximated and Detailed Wavelet Transform coefficients in the main memory. Also, coefficients can be stored in an 8\,bit format to compress information in the main memory. The accelerator is programmed at the beginning with the number of samples to acquire and the output vector pointers. The eFPGA autonomously loops over SPI transactions and stores to the main memory, either the raw data or the Approximated and Detailed coefficients of the HDWT. When all the $N$ data have been stored into the memory, an interrupt notifies the core at the end of the acquisition.

Moreover, a second function has been mapped to the custom SPI peripheral, namely, to extract 4\,bits local binary patterns from a stream of data coming from sensors, as an algorithmic approach presented in \cite{Burrello}. In this case, for each data acquired, the eFPGA reuses the subtractor instantiated for the HDWT to compare the last two samples. If the last sample is greater than the previous one, it stores 1 in a 4\,bit shift register, otherwise 0. The accelerator stores into memory a 16\,bit value every four samples, each representing four single sample overlapping windows. The core takes 8 cycles for each tuple approximate-detail coefficient to compute the HDWT, whereas it takes 16 cycles for the local binary pattern. The eFPGA instead computes the features during the acquisition of the signal from SPI without adding latency overheads.

The design utilizes \FPGAAREAspipulpino{} of the available SLCs, and it uses a memory interface port, the \ac{APB} interface, four GPIOs (3 output pins and 1 input pin), and it generates one event.

\subsection{Custom \io{} interface}

IoT devices are often connected to custom peripherals that need more control pins that the usual peripherals as SPI, UART, I2C, I2S, etc. In this case, off-chip FPGAs are selected to implement the control part of the custom peripheral on one side and to communicate with the \ac{MCU} with a standard protocol (e.g., SPI) to the other side. An example of a custom peripheral is a neuromorphic vision sensor \cite{Inivation} or event-based audition sensors \cite{Liu}. Another example where FPGAs are used to control and transfer data are bridges for off-chip accelerators, for example, \cite{Conti18}, or \cite{Lee}. In this context, to illustrate the flexibility of the \ac{MCU}+eFPGA combination, a controller for the systolic Long short-term memory Recurrent Neural Network (LSTM-RNN) accelerator presented in \cite{Conti18} has been implemented in the eFPGA. The LSTM-RNN accelerator is made of four chips implemented in UMCL 65\,nm technology, and it is used to classify phonemes in real-time. The eFPGA uses 36\,GPIOs to interact with the accelerator using a custom interface.

\begin{figure}
    \centering
\centerline{\includegraphics[width=\columnwidth]{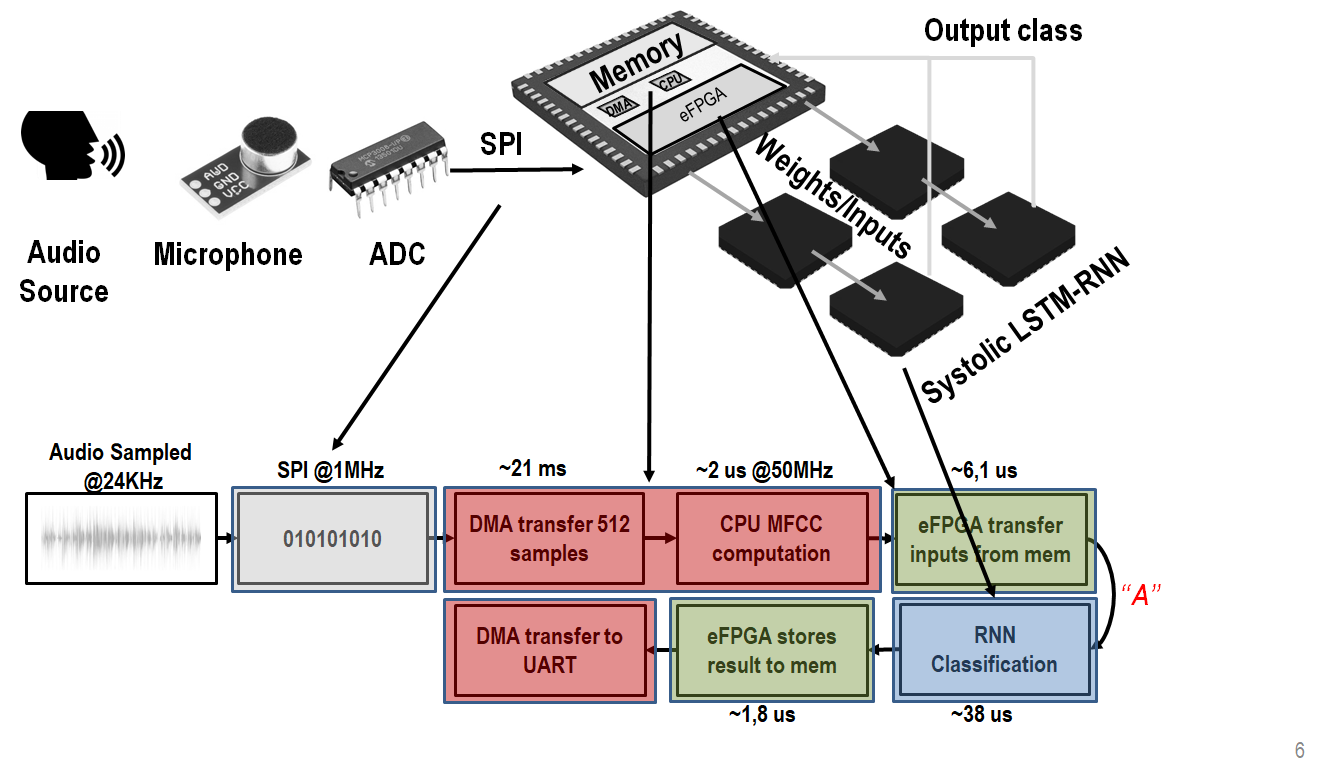}}
    \caption{Example of an application where the proposed design is driving custom protocol off-chip accelerators. Data coming from microphones are first pre-processed by the \ac{MCU}, then sent to the off-chip accelerator via eFPGA for classification.}
    \label{fig:muntaniala}
    \vspace{-4mm}
\end{figure}

In the first phase, the eFPGA sends the weights of the RNN-model into the four chips. Then, for every sample acquired by the \ac{MCU} \io{} subsystem, the CPU extracts the Mel-Frequency Cepstral Coefficients (MFCCs). In parallel, the eFPGA autonomously fetches the coefficients from the main memory of the \ac{MCU} and sends them to the off-chip accelerator. Once the inference on the accelerator has been computed, the result is sent back to the eFPGA, which stores it to the main memory of the \ac{MCU} and finally notifies the core with an interrupt. 
\figref{fig:muntaniala} shows the data flow from the microphone to the accelerator and back to the \ac{MCU}. The utilization of the eFPGA is only \FPGAAREAmuntaniala{}. Managing 36\,GPIOs through \ac{MCU} firmware (of which one is actually the clock of the off-chip accelerator) would require the core to run at higher frequency than the eFPGA due to the sequential nature of software. In this example the external accelerator is running at 80\,MHz. This means that in the best case, the CPU should be able to perform \textasciitilde{}7 operations in 12.5\,ns, which requires 560\,MHz, and 2.5$\times$ higher energy consumption than the eFPGA based solution.


\subsection{CPU subsystem accelerator}
\label{sub:cpuacc}

In the context of on-the-edge computation, accelerators are used to increase performance and the energy efficiency of such devices \cite{7805364}. For pattern recognition tasks in the visual domain, deep quantized neural networks are an attractive model due to its limited memory and computational requirements \cite{hubara2017quantized}. In extreme cases, single-bit representation for weights and data is chosen to minimize the memory footprint and the computational resources, as it requires simple operations as logic XOR rather than multiplications to compute convolutions. Such neural networks are called Binary Neural Networks (BNN) \cite{courbariaux2015binaryconnect, rastegari2016xnor}. The eFPGA has sufficient resources to allow these accelerators to be implemented, freeing the core for other computing tasks.

The BNN accelerator designed for this scope has four interfaces towards the main memory to maximize the bandwidth, and it is a simplified version of the accelerator presented in \cite{ContiBNN}. It assumes that input layers and filters are organized as a 3D array (number of filters $\times$ rows $\times$ columns) of integers, where each integer represents a 32 one-bit channels. The accelerator is implemented to operate on two 3$\times$3 windows with eight filters f0, ..., f7 in parallel to simplify the controlling part, but this is not a limiting factor for the use-case under study. The accelerator is programmed via the APB interface by the core with the output, input and filter layer pointers, the number of rows and columns of the input layer, and with the \textsc{START} command. The eFPGA starts by fetching two 32\,bit input elements, then four 32\,bit elements are fetched in parallel twice to acquire the eight filter elements.

The eFPGA performs the \textsc{XOR} function between the inputs and the eight filters, accumulates all the single-bit partial results. The sixteen 3$\times$3 convolution results are then compared with a programmed threshold to compute the activation functions. The accelerator autonomously iterates over the input rows and columns; then, it sends an interrupt to the core to signal the end of the computation. During this period, the core can wait for the accelerator to finish in \textsc{IDLE} mode to save power or deal with other tasks in parallel (for example scheduling the next \io{} tasks, elaborating previously filtered data, etc.). The design occupies \FPGAAREAbnnacc{} of the SLCs available, and it uses 4 memory interfaces, the \ac{APB} port, and it generates 1 event. The application consumes 12.5\,mW (eFPGA+MCU), and it runs in 371\,\textmu s at 125\,MHz. Although the core implements custom instructions to speed up such kernels (as the pop count instruction), and it can run faster (600\,MHz against 125\,MHz), to implement the same function the CPU consumes 15\,mW, and it runs in 675\,\textmu s, with an energy efficiency 2.2$\times$ lower than the eFPGA.

As a second CPU accelerator, a cyclic redundancy check (CRC) accelerator has been implemented in the eFPGA to ensure data integrity and error correction \cite{7558141}. Such an accelerator uses the \io{} DMA interface to leverage the linear address generator already present in the $\mu$DMA and thus saving resources in the eFPGA. The CPU programs the $\mu$DMA to fetch data from the L2 memory and transmits them to the eFPGA accelerator, which calculates the CRC value. The accelerator has a register to know the number of data to process, whereas the read- and write-pointers are written in the $\mu$DMA configuration registers.  This low area accelerator consumes only \FPGAAREAcrcacc{} of the SLCs available, and it only uses 1 interface towards the \textmu DMA with configuration, TX/RX ports. The application consumes only 7.5\,mW (eFPGA+MCU), and it runs in 3.7\,\textmu s at 193\,MHz for 1024 byte data. The CPU consumes 15\,mW, and it runs in 78\,\textmu s, with an energy efficiency 42.2$\times$ less than the eFPGA.
\textcolor{black}{To compare the performance of the proposed eFPGA-based system with respect to the Microsemi PolarFire IoT gateway-class FPGA SoC \cite{polarfire}, the power estimator from Microsemi has been used. Results show a power consumption of 111\,mW, 14.8$\times$ higher than our work. The estimation has been performed setting the same frequency, number of LUTs and flip-flops.}

%
%

\tabref{tab:use_case} shows the number of GPIOs,  number of flip-flops (FF), and LUTs required by each use case. Power figures (expressed in mW) correspond to the system when the eFPGA runs, and the CPU waits for the result, whereas the final  column shows the energy gained by running the accelerator on the eFPGA rather than software. In the Custom \io{} example, the SW could not handle the protocol at the speed required, for that example eFPGA was the only viable solution.

Basic interfaces like I2C and UART have been implemented on the eFPGA using the DMA interface with about 5\% of eFPGA resources, and a more complex parallel camera interface with full DMA support implementation uses only 12\% of available eFPGA resources.


\section*{Comparison with SoA}

\tabref{tab:soa_table} shows a comparison with various chips reported in the literature. The table includes heterogeneous reconfigurable systems composed of \ac{MCU} and eFPGA, an embedded domain FPGA SoC, and an advanced low-power \ac{MCU}s in 28\,nm \ac{FDSOI}. The standalone \ac{MCU} \cite{Bol19} has a 4$\times$ smaller power density (\textmu W/MHz). However, our \ac{MCU} features 8x larger memory capacity and significantly larger peak performance as well: 7.5$\times$ higher maximum frequency, 3.19 vs. 2.33\,Coremark/MHz, and almost 6$\times$ better performance in near-sensor processing workloads when compared to the ARM Cortex M0 processor used in \cite{Bol19}. Hence, our energy efficiency on the targeted application domain is 1.5$\times$ better.

The advanced \ac{MCU}+eFPGA system presented in \cite{Whatmough} is a high-performance class system implemented in  25\,mm\textsuperscript{2}, where a bigger eFPGA (6$\times$ higher leakage power), two application class 64\,bit cores, a quad-core cluster accelerator, and 12$\times$ bigger memory are used (including caches). The eFPGA offers 80 MACs blocks, more LUTs, and eFPGA flip-flops, and provides remarkable energy efficiency of 312\,GOPS/W. Thanks to the abundance of DSP blocks in the FPGA fabric. However, this system is meant to be used in high-performance applications consuming higher dynamic and leakage power not suitable for IoT applications. On the other hand, Arnold, although achieving a lower peak efficiency, is in a power range suitable for IoT applications (below hundreds of mW). Moreover, the reverse body biasing applied to the FPGA fabric can reduce leakage power to a value as low as 20.5\,\textmu W, more than two orders of magnitude better than \cite{Whatmough}. The Microsemi SmartFusion2 SoC \cite{SmartFusion} used in \cite{Fournaris} is built in 65\,nm. The whole system can run up to 160\,MHz (> 3.75$\times$ slower than the proposed work), and it achieves 21$\times$ higher power density. The works of Borgatti \cite{Borgatti} and Lodi \cite{Lodi} exploit embedded reconfigurable datapaths to accelerate DSP patterns of signal processing applications, achieving remarkable performance and operating frequency despite the old nodes used for implementation. With respect to these works and the other heterogeneous \ac{MCU}+eFPGA systems of the same class \cite{Renzini, Lodi, Borgatti}, the proposed SoC has more than 2.9$\times$ better efficiency, more than 3.4$\times$ better performance, and more than 2.2$\times$ larger capacity. Moreover, this is the first design offering flexible connections enabling reconfigurable peripherals, \io{} accelerators, shared-memory accelerators, and supporting state-retentive deep sleep based on reverse body bias, paving the way for flexible fully programmable IoT end-nodes.

\section{Conclusion}

In this paper, we presented \textit{Arnold}; a RISC-V based MCU extended with an embedded FPGA for flexible power-constraints energy-efficient IoT devices. The system has built-in \ac{GF22FDX}, it occupies 9\,mm\textsuperscript{2}, and it leverages body bias to tune performance-power trades off. The eFPGA is a 32$\times$32 array macro provided by QuickLogic connected to the rest of the system through four parallel memory interfaces (128\,bit per transaction); a TX/RX \io{} DMA interface; sixteen events to interact with the CPU; GPIOs; and APB. The paper shows how the eFPGA can be used to extend and accelerate the SoC peripheral subsystem, as well as a CPU accelerator. The eFPGA has more than 6K LUTs and 4K flip-flops, enough to implement standard and custom peripherals used in the IoT domain and simple accelerators to enhance the energy efficiency of the SoC. It achieves 46.83\,\textmu W/MHz, top in class in the mW domain of IoT devices. The CPU runs up to 600\,MHz (620 with FBB), more than 7$\times$ faster than the best energy efficient MCU. Leakage power of the whole system can be as low as 552\,\textmu W when the MCU runs at 0.5\,V, and the eFPGA is kept in state retentive deep-sleep via RBB. The paper shows that integrating an eFPGA in an MCU in \ac{GF22FDX} gives to IoT devices the high versatility needed for extended product life and shorter time-to-market, still without waiving performance, power and energy efficiency.

\section*{Acknowledgment}
This work has received funding from the European Union’s Horizon 2020 research and innovation program under grant agreement No 732631, project ``OPRECOMP''.

\bibliographystyle{IEEEtran}
\bibliography{bibl.bib}
\bibliographystyle{IEEEtran}
\vspace{-1.2cm}
\begin{IEEEbiography}[{\includegraphics[width=1in,height=1.25in,clip,keepaspectratio]{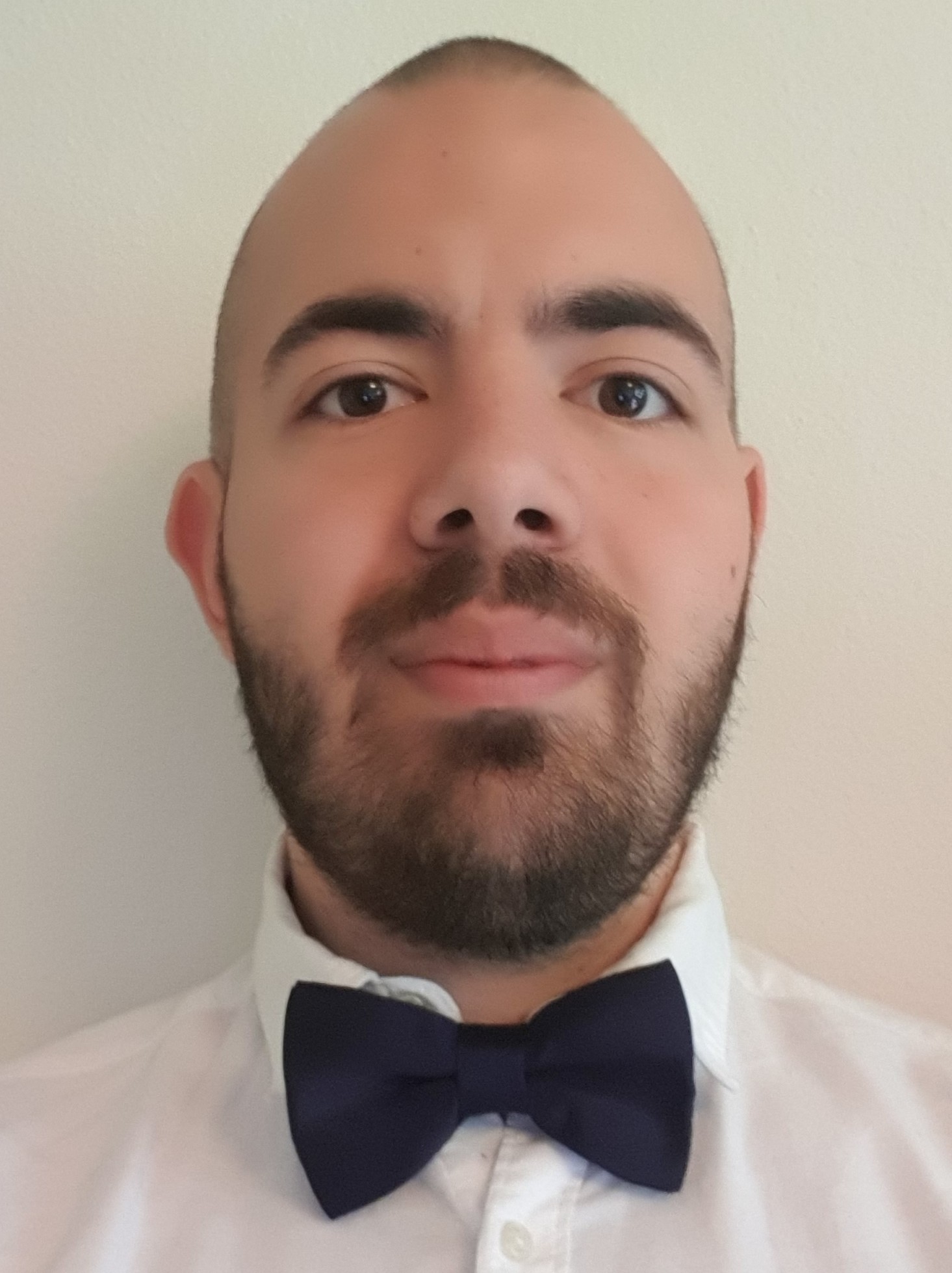}}]{Pasquale Davide Schiavone}
received his B.Sc. (2013) and M.Sc. (2016) in computer engineering from Polytechnic of Turin. In 2016 he has started his Ph.D. studies at the Integrated Systems Laboratory, ETH Zurich. In 2018, he has been Ph.D visiting student in the Centre for Bio-Inspired Technology, Imperial College London. His research interests include datapath blocks design, low-power microprocessors in multi-core systems and deep-learning architectures for energy-efficient systems.
\end{IEEEbiography}
\vspace{-1.2cm}
\begin{IEEEbiography}[{\includegraphics[width=1in,height=1.25in,clip,keepaspectratio]{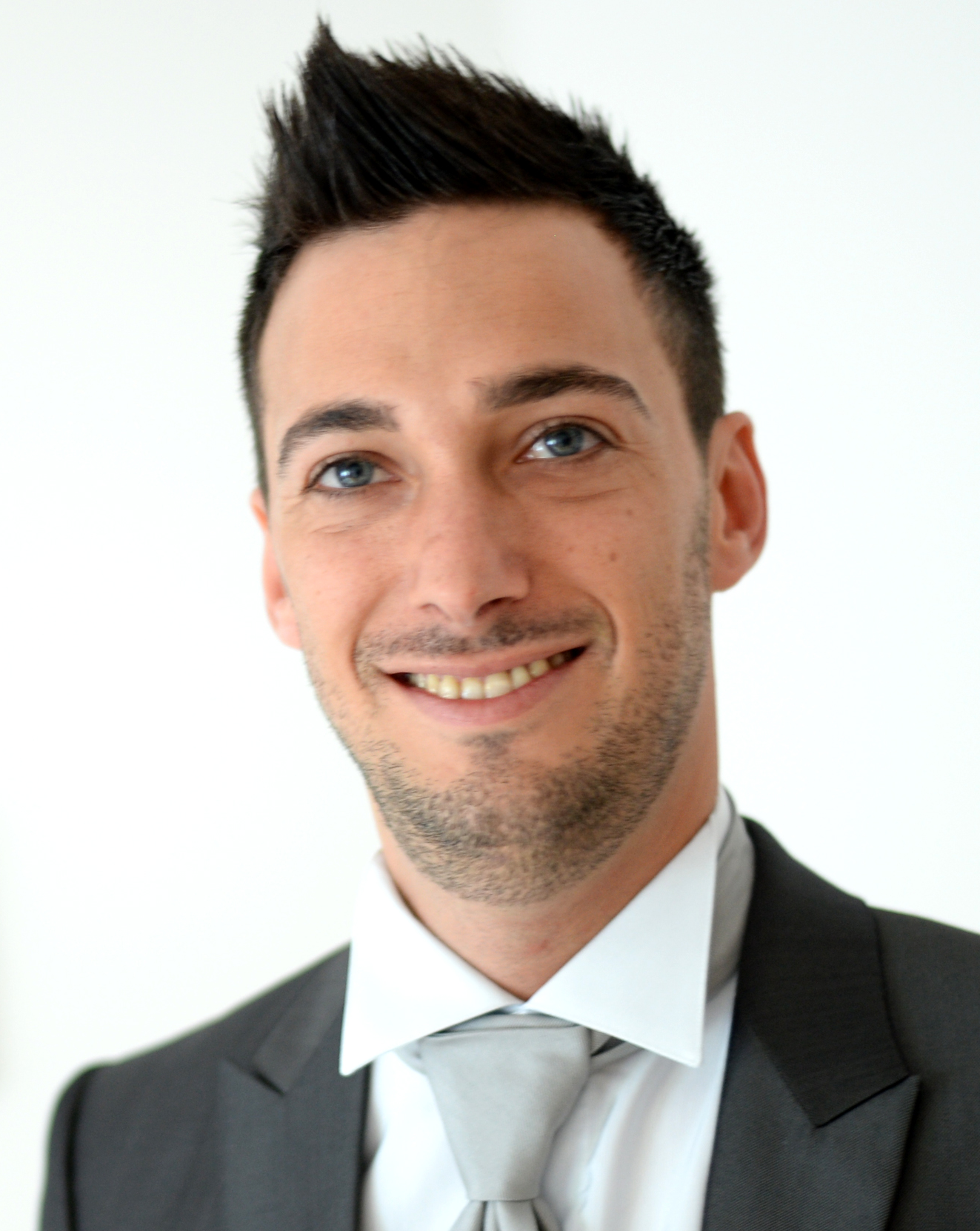}}]{Davide Rossi}, received the PhD from the University of Bologna, Italy, in 2012. He has been a post doc researcher in the Department of Electrical, Electronic and Information Engineering “Guglielmo Marconi” at the University of Bologna since 2015, where he currently holds an assistant professor position. His research interests focus on energy efficient digital architectures in the domain of heterogeneous and reconfigurable multi and many-core systems on a chip. This includes architectures, design implementation strategies, and runtime support to address performance, energy efficiency, and reliability issues of both high end embedded platforms and ultra-low-power computing platforms targeting the IoT domain. In these fields he has published more than 100 papers in international peer-reviewed conferences and journals. He is recipient of Donald O. Pederson Best Paper Award 2018.
\end{IEEEbiography}
\vspace{-1.2cm}
\begin{IEEEbiography}[{\includegraphics[width=1in,height=1.25in,clip,keepaspectratio]{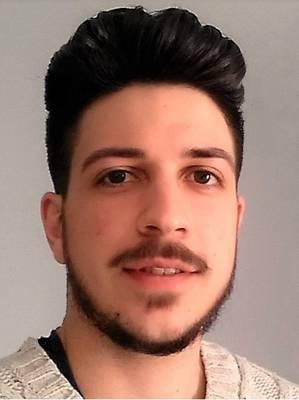}}]{Alfio Di Mauro}
received his M.Sc. degree in Electronic Engineering from the Electronics and Telecommunications Department (DET) of Politecnico di Torino in 2016.
In January 2017, he started to work as researcher assistant in the Integrated System Laboratory (IIS) of the Swiss Federal Institute of Technology of Zurich, in the group led by Prof. Luca Benini. 
Since September 2017, he is pursuing the PhD in electrical engineering in the same Laboratory. His research is mainly focused on the design of digital Ultra-Low Power (ULP) System-on-Chip (SoC) for Event-Driven edge computing.
\end{IEEEbiography}
\vspace{-1.2cm}
\begin{IEEEbiography}[{\includegraphics[width=1in,height=1.25in,clip,keepaspectratio]{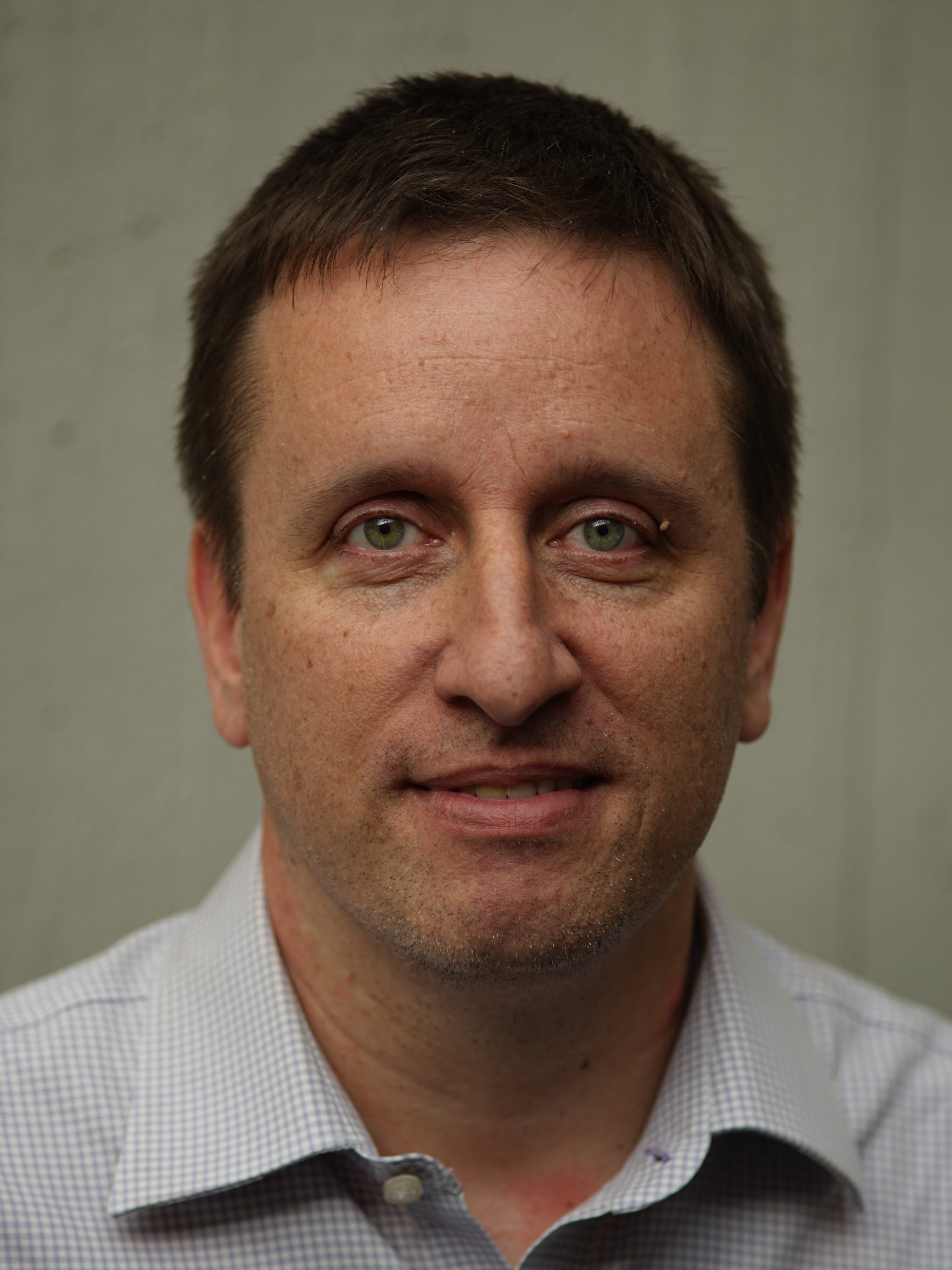}}]{Frank K. G\"urkaynak}
has obtained his B.Sc. and M.Sc, in electrical engineering from the Istanbul Technical University, and his Ph.D. in electrical engineering from ETH Z\"urich in 2006. He is currently working as a senior researcher at the Integrated Systems Laboratory of ETH Z\"urich. His research interests include digital low-power design and cryptographic hardware.
\end{IEEEbiography}
\vspace{-1.2cm}
\begin{IEEEbiography}[{\includegraphics[width=1in,height=1.25in,clip,keepaspectratio]{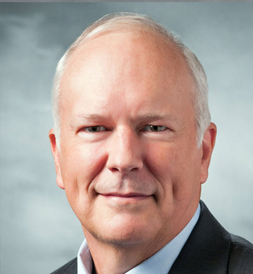}}]{Timothy Saxe}
(Ph.D.) joined QuickLogic in May 2001. Dr. Saxe has served as our Senior Vice President of Engineering and Chief Technology Officer since August 2016 and Senior Vice President and Chief Technology Officer since November 2008. Previously, Dr. Saxe has held a variety of executive leadership positions in QuickLogic including Vice President of Engineering and Vice President of Software Engineering. Dr. Saxe was Vice President of Flash Engineering at Actel Corporation, a semiconductor manufacturing company, from November 2000 to February 2001. Dr. Saxe joined GateField Corporation, a design verification tools and services company formerly known as Zycad, in June 1983 and was a founder of their semiconductor manufacturing division in 1993. Dr. Saxe became GateField’s Chief Executive Officer in February 1999 and served in that capacity until Actel Corporation acquired GateField in November 2000. Dr. Saxe holds a B.S.E.E. degree from North Carolina State University, and an M.S.E.E. degree and a Ph.D. in Electrical Engineering from Stanford University..
\end{IEEEbiography}
\vspace{-1.2cm}\begin{IEEEbiography}[{\includegraphics[width=1in,height=1.25in,clip,keepaspectratio]{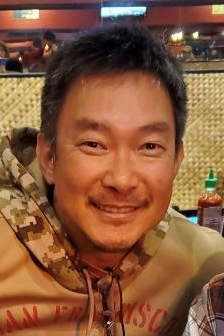}}]{Mao Wang} is a Sr. Director of Product at QuickLogic, with the mission of democratizing embedded FPGA into every SoC.
He holds a B.S. degree en Electrical Engineering and a M.S. degree in Engineering Management from Santa Clara University.

\end{IEEEbiography}
\vspace{-1.2cm}\begin{IEEEbiography}[{\includegraphics[width=1in,height=1.25in,clip,keepaspectratio]{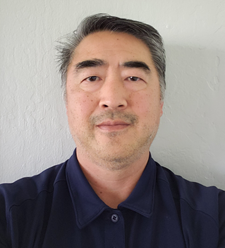}}]{Ket Chong Yap}
joined QuickLogic in September 1999, actively participating in QuickLogic FPGA product development.  Prior to joining QuickLogic, Mr. Yap was with EXEL Microelectronics as Quality Assurance Engineer from 1990 to 1991, Product/Test Engineer from 1992 to 1994, and Design Engineer from 1995 to 1996, working on EEPROM technology. Mr. Yap was also involved with Programmable Microelectronics Corporation from 1997 to 1998, working on FLASH memory. Mr. Yap hold a B.S. degree in Electrical Engineering from the Iowa State University, Ames.
\end{IEEEbiography}
\vspace{-1.2cm}
\begin{IEEEbiography}[{\includegraphics[width=1in,height=1.25in,clip,keepaspectratio]{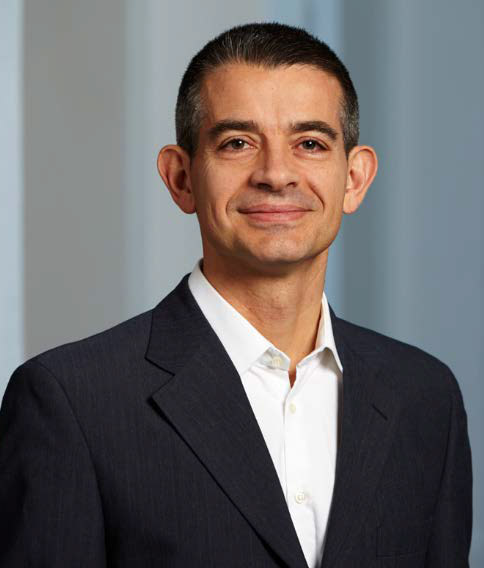}}]{Luca Benini}
(F’07) received the Ph.D. degree in electrical engineering from Stanford University, Stanford, CA, USA, in 1997.
He has served as the Chief Architect of the Platform 2012/STHORM Project with STMicroelectronics, Grenoble, France, from 2009 to 2013. He held visiting/consulting positions with \'Ecole Polytechnique F\'ed\'erale de Lausanne, Stanford University, and IMEC. He is currently a Full Professor with the University of Bologna, Bologna, Italy. He has authored over 700 papers in peer-reviewed international journals and conferences, four books, and several book chapters. His current research interests include energy-efficient system design and multicore system-on-chip design.
Dr. Benini is a member of Academia Europaea. He is currently the Chair of Digital Circuits and Systems with ETH Z\"urich, Z\"urich, Switzerland.
\end{IEEEbiography}

\end{document}